\documentclass[a4paper,12pt]{article}
\pdfoutput=1
\usepackage{epsfig}
\usepackage{amssymb}
\usepackage{amsfonts}
\usepackage{amsmath}
\usepackage{euscript}
\usepackage{verbatim}
\usepackage{latexsym}
\usepackage{graphicx}
\usepackage{caption}
\usepackage{float}
\usepackage{subcaption}

\usepackage{listings}

\usepackage{booktabs}

\newif\ifdtup

\catcode`\@=11

\@addtoreset{equation}{section}

\def\@normalsize{\@setsize\normalsize{15pt}\xiipt\@xiipt
\abovedisplayskip 14pt plus3pt minus3pt%
\belowdisplayskip \abovedisplayskip
\abovedisplayshortskip \z@ plus3pt%
\belowdisplayshortskip 7pt plus3.5pt minus0pt}

\def\small{\@setsize\small{13.6pt}\xipt\@xipt
\abovedisplayskip 13pt plus3pt minus3pt%
\belowdisplayskip \abovedisplayskip
\abovedisplayshortskip \z@ plus3pt%
\belowdisplayshortskip 7pt plus3.5pt minus0pt
\def\@listi{\parsep 4.5pt plus 2pt minus 1pt
     \itemsep \parsep
     \topsep 9pt plus 3pt minus 3pt}}

\relax

\catcode`@=12

\catcode`\@=11

\def\section{\@startsection{section}{1}{\z@}{3.5ex plus 1ex minus
   .2ex}{2.3ex plus .2ex}{\large\bf}}

\def\SymBoxes#1#2#3#4{\newdimen\un@t \un@t#3%
\raisebox{#1}{\rule{#2\un@t}{#4}\hskip-#2\un@t
\@tempdimb\un@t \advance\@tempdimb by-#4\@tempcntb#2\relax%
\@whilenum{\@tempcntb>0}\do{
\rule{#4}{\un@t}\hskip\@tempdimb \advance\@tempcntb by\m@ne}%
\hskip-#2\un@t \rule[\un@t]{#2\un@t}{#4}%
\rule[\un@t]{#4}{#4}\hskip-#4
\rule{#4}{\un@t}}\hskip-#4}                

\begin{document}

\newcommand{\beq}{\begin{equation}}
\newcommand{\eeq}{\end{equation}}
\newcommand{\bea}{\begin{eqnarray}}
\newcommand{\eea}{\end{eqnarray}}
\newcommand{\beas}{\begin{eqnarray*}}
\newcommand{\eeas}{\end{eqnarray*}}
\newcommand{\defi}{\stackrel{\rm def}{=}}
\newcommand{\non}{\nonumber}
\newcommand{\bquo}{\begin{quote}}
\newcommand{\enqu}{\end{quote}}
\renewcommand{\(}{\begin{equation}}
\renewcommand{\)}{\end{equation}}
\def \eqn#1#2{\begin{equation}#2\label{#1}\end{equation}}

\def\e{\epsilon}
\def\IZ{{\mathbb Z}}
\def\IR{{\mathbb R}}
\def\IC{{\mathbb C}}
\def\IQ{{\mathbb Q}}
\def\de{\partial}
\def\Tr{ \hbox{\rm Tr}}
\def\H{ \hbox{\rm H}}
\def\HE{ \hbox{$\rm H^{even}$}}
\def\HO{ \hbox{$\rm H^{odd}$}}
\def\K{ \hbox{\rm K}}
\def\Im{ \hbox{\rm Im}}
\def\Ker{ \hbox{\rm Ker}}
\def\const{\hbox {\rm const.}}
\def\o{\over}
\def\im{\hbox{\rm Im}}
\def\re{\hbox{\rm Re}}
\def\bra{\langle}\def\ket{\rangle}
\def\Arg{\hbox {\rm Arg}}
\def\Re{\hbox {\rm Re}}
\def\Im{\hbox {\rm Im}}
\def\exo{\hbox {\rm exp}}
\def\diag{\hbox{\rm diag}}
\def\longvert{{\rule[-2mm]{0.1mm}{7mm}}\,}
\def\a{\alpha}
\def\dag{{}^{\dagger}}
\def\tq{{\widetilde q}}
\def\p{{}^{\prime}}
\def\W{W}
\def\N{{\cal N}}
\def\hsp{,\hspace{.7cm}}

\def\br{\nonumber}
\def\IZ{{\mathbb Z}}
\def\IR{{\mathbb R}}
\def\IC{{\mathbb C}}
\def\IQ{{\mathbb Q}}
\def\IP{{\mathbb P}}
\def \eqn#1#2{\begin{equation}#2\label{#1}\end{equation}}

\newcommand{\C}{\ensuremath{\mathbb C}}
\newcommand{\Z}{\ensuremath{\mathbb Z}}
\newcommand{\R}{\ensuremath{\mathbb R}}
\newcommand{\rp}{\ensuremath{\mathbb {RP}}}
\newcommand{\cp}{\ensuremath{\mathbb {CP}}}
\newcommand{\vac}{\ensuremath{|0\rangle}}
\newcommand{\vact}{\ensuremath{|00\rangle}                    }
\newcommand{\oc}{\ensuremath{\overline{c}}}
\newcommand{\psizero}{\psi_{0}}
\newcommand{\phizero}{\phi_{0}}
\newcommand{\hzero}{h_{0}}
\newcommand{\psiin}{\psi_{\rh}}
\newcommand{\phiin}{\phi_{\rh}}
\newcommand{\hin}{h_{\rh}}
\newcommand{\rh}{r_{h}}
\newcommand{\rb}{r_{b}}
\newcommand{\psibnd}{\psi_{0}^{b}}
\newcommand{\psibndp}{\psi_{1}^{b}}
\newcommand{\phibnd}{\phi_{0}^{b}}
\newcommand{\phibndp}{\phi_{1}^{b}}
\newcommand{\gbnd}{g_{0}^{b}}
\newcommand{\hbnd}{h_{0}^{b}}
\newcommand{\zh}{z_{h}}
\newcommand{\zb}{z_{b}}
\newcommand{\man}{\mathcal{M}}
\newcommand{\hbr}{\bar{h}}
\newcommand{\tbr}{\bar{t}}

\begin{titlepage}

\def\thefootnote{\fnsymbol{footnote}}

\begin{center}
{\bf
{\large Virasoro Blocks and Trouble at the Euclidean Horizon} \\
}
\end{center}

\begin{center}
Aaditya Datar$^{a,b}$\footnote{\texttt{aadityadatar.physics@gmail.com}}, Chethan Krishnan$^b$\footnote{\texttt{chethan.krishnan.physics@gmail.com}} \  

\end{center}

\renewcommand{\thefootnote}{\arabic{footnote}}

\begin{center}
$^a$ {Indian Institute of Science Education and Research, \\
Homi Bhabha Rd, Pune 411008, India}\\
\vspace{0.2in}

$^b$ {Center for High Energy Physics,\\
Indian Institute of Science, Bangalore 560012, India}\\

\end{center}
\vspace{-0.15in}
\noindent
\begin{center} {\bf Abstract} \end{center}
In the semi-classical ($c \rightarrow \infty$) limit, 4-point HLLH correlators in 2D CFTs exhibit periodic Euclidean singularities. Periodic singularities in Euclidean time are a general feature of thermal correlators, even at weak coupling. Therefore, the bulk significance of this observation (in particular, the role of the horizon) is somewhat obscure. Explicit numerical computations of finite-$c$ Virasoro blocks furthermore suggest that their departure from semi-classical blocks may begin already at half the period. In this paper, we provide a bulk understanding of these facts and clarify the role of the horizon. We present a bulk geodesic Witten diagram calculation of semi-classical Virasoro blocks in coordinates that are naturally adapted to the BTZ black hole. This allows a bulk geometric interpretation for boundary time separation. In this language, half of a thermal time period is the boundary timescale at which the light operator geodesic straddles the Euclidean horizon, capturing both the role of the horizon and the associated timescale. This timescale arises in a calculation that does {\em not} involve a periodic thermal circle on the bulk or the boundary. 




\vspace{1.6 cm}
\vfill

\end{titlepage}

\setcounter{footnote}{0}

\tableofcontents

\section{A Euclidean Information Paradox?}

In this paper, we will be interested 4-point scalar correlators in large-$c$ holographic 2D CFTs of the type 
\bea
\langle O_H O_L O_L O_H \rangle.
\eea
These HLLH correlators have turned out to be instructive in understanding some aspects of the information paradox. $O_L$ here is a light operator, its scaling dimension $h_L$ satisfies $h_L/c \rightarrow 0$ in the large-$c$ limit. For the heavy operator $O_H$, the ratio $h_H/c$ stays finite as $c$ is sent to infinity. Equivalently, $O_L$ captures a probe field in the bulk, whereas $O_H$ is a black hole microstate. In the large-$c$ limit, we expect the above correlator to be indistinguishable from that of a light probe in a black hole background -- therefore it should (see e.g., \cite{Catalan}) exhibit information loss. Concretely, 
\bea
\langle O_H O_L O_L O_H \rangle = \langle O_H| O_L O_L | O_H\rangle \xrightarrow[c \rightarrow \infty]{}
\langle O_L O_L \rangle_{T_H} \label{mainformula}
\eea
where we have used state-operator correspondence in the first equality. The final expression on the right is the thermal correlator, and $T_H=\frac{1}{2 \pi}\sqrt{24 h_H/c-1}$ is well-defined in the large-$c$ limit and equals the Hawking temperature of the black hole. At finite-$c$ on the other hand, the left hand side is a perfectly well-defined pure state correlator in a unitary CFT. Therefore understanding HLLH correlators at finite-$c$, especially from the bulk, should be a valuable exercise in understanding information loss. The large-$c$ limit in which the scaling dimensions are scaled in the above manner is what we will understand by the ``semi-classical limit" in this note.

Information paradox is a {\em bulk} problem in AdS/CFT, because the CFT is manifestly unitary\footnote{It may be worth understanding {\em what} aspect of the CFT is causing the dual bulk description to be apparently non-unitary. It seems plausible that this has to do with the huge number of states (scaling exponentially in $N$) that are present in the black hole energy sliver.}. What is lacking, is a clear understanding of the mechanism for information restoration in the bulk, because the bulk theory is not {\em manifestly} unitary. So if one wants to make progress towards a resolution of the information paradox, it will be useful to study the above 4-point function, from the bulk. There are two key tools we will use in this paper, to approach this task -- (a) Virasoro blocks, and (b) geodesic Witten diagrams to compute these blocks semi-classically from the bulk. We briefly review what a Virasoro block is in Appendix \ref{AppVirasoro}, and will outline the necessary aspects of geodesic Witten diagram prescriptions in the next section.

Black hole information paradox is generally viewed as a {\em Lorentzian} problem, and therefore formula \eqref{mainformula} is most directly applicable, in real time. However, the fact that thermal two-point functions are periodic in Euclidean time, together with the presence of coincidence singularities when the light operators approach each other, suggests that the thermal correlator has an infinite number of periodic singularities in imaginary time. Such singularities must necessarily be absent on the left-hand side of \eqref{mainformula}, because only coincidence singularities are allowed in pure state correlators in a Euclidean CFT. This can be viewed as a type of information paradox \cite{KapInfo}. It should be emphasized however, that such periodic singularities are a general feature of thermal systems -- they are not exclusive to black holes or holographic CFTs. More specifically, the statement about the existence of periodic singularities is independent of the existence of a bulk horizon. A closely related fact is that the loss of fine-grained information due to thermalization happens in weakly coupled systems as well, and is usually not viewed as a paradox\footnote{This statement requires a few qualifiers. In genuinely free theories with a Fock space, there is no known mechanism for microcanonical thermalization -- the system is integrable. Eigenstate thermalization (ETH) \cite{Deutsch, Srednicki} does not help us there. This means that in order for a Planckian gas to thermalize, we have to assume weak self-interactions, or assume interactions with the phonons on the wall of the box (the latter have to be assumed to not disturb the microcanonical assumption too much). See \cite{Moore}  for a discussion of weakly coupled thermalization in non-Abelian plasmas. When considering canonical thermalization, such subtleties are usually unimportant, because one assumes that the system has coupling to a thermal bath.}. One of our goals in this paper is to show how the horizon geometrizes the information loss discussed in \cite{KapInfo}, when it comes to holographic 2D CFTs. This allows a specific sense in which we can claim that the horizon is ``responsible'' for Euclidean information loss as well.

The Virasoro block is interesting for our purposes because some of the features of information loss in the 4-point function are reflected faithfully in the block in its semi-classical limit. In particular, semi-classical Virasoro blocks exhibit the periodic singularities discussed above \cite{KapInfo}. So our goal in this paper will be to compute the semi-classical block using bulk geodesic Witten diagrams so that we can identify the precise role of the horizon. We will see that information loss can be associated to bulk geodesics probing the horizon radius in the geodesic Witten diagram prescription. 

In order to accomplish this, we will develop a bulk geodesic Witten prescription that is different from the one previously used in \cite{Kraus}. The prescription of \cite{Kraus} did the construction using the conical defect metric and then used analytic continuation beyond the BTZ threshold to reach the loss-y Virasoro blocks. We discuss this in greater detail in the next section, but here we will simply observe that such an approach cannot be used to directly connect with bulk physics above the BTZ threshold. Instead we will develop a prescription that is directly applicable in the BTZ metic. Using this prescription, we will be able to argue that information loss is directly related to the bulk geodesic reaching down to the horizon scale. 

What does this observation mean? The natural suggestion is that new physics at the horizon scale will be relevant for resolving information loss even in the Euclidean regime. Similar ideas have appeared before in the context of string theory in the the Euclidean cigar geometry in both 1+1 \cite{Giveon, Itzhaki, Yoav} and higher \cite{Sarthak} dimensions. The mechanism there is that the geometry needs modification at the tip of the cigar (more concretely, a puncture emerges). It is noteworthy that in our calculation also, the bulk geometry associated to the computation of the Virasoro block ends abruptly\footnote{There is no periodic identification and no cigar.} at the horizon radius. A further curiosity we will note is that the geodesics that we find that straddle the horizon radius in our block calculations (and are therefore able to stretch without much cost), have analogies to the winding strings that condense and produce the puncture in the string theory story\footnote{Note that the bulk geometry associated to a Virasoro block does not have a periodic identification -- if it did, the emergence of the periodic singularities would be less interesting.}. This suggests that it may be interesting to generalize the geodesic Witten diagram calculation of the block/correlator that we discuss here, to a worldsheet Witten diagram calculation of some kind. 

In what follows, we start with a brief review of the conical defect prescription for geodesic Witten diagrams \cite{Kraus}. Then we turn to the BTZ version of the same calculation. Even though in the end both yield the semi-classical Virasoro block, the philosophy and details of the calculation are different - so we discuss it in some detail. After explaining the approach, we show that it reproduces the correct semi-classical block. The advantage of the approach is that the loss-y regime of the block has a geometric interpretation --  it is directly associated to the bulk geodesic probing the horizon radius. We use this to explain various aspects of the physics, and some features of finite-$c$ Euclidean Virasoro blocks numerically computed in \cite{KapN}. In a technical Appendix \ref{AppFourier}, we present an explicit Fourier transformation relating position and momentum space 2-point functions in the BTZ black hole. We have placed this in an Appendix because it is technically somewhat distinct from the rest of the paper, but conceptually it may be a necessary first step in probing black hole horizons using correlators/Virasoro blocks and going beyond the present paper. 



\section{Geodesic Witten Diagrams for Semi-Classical Virasoro Blocks}

The geodesic Witten diagram prescription for a semi-classical Virasoro block in AdS$_3$/CFT$_2$ is given by
\bea\label{3a}
    \mathcal{W}_{2h,0}(x_i)\hspace{-0.1cm} 
    \equiv \hspace{-0.2cm}\int d\lambda\;G_{b\partial}(x_1, y(\lambda))G_{b\partial}(x_2, y(\lambda))\int d\lambda'\;G_{b\partial}(x_3, y(\lambda'))G_{b\partial}(x_4, y(\lambda')) \ G_{bb}(y(\lambda), y(\lambda'); 2h) \nonumber \\
    \label{GWD}
\eea
The diagram that one usually associates to this formula can be seen (for example) in Figure 2 of \cite{Kraus}. The prescription is loosely the same as that for computing global conformal blocks in the original geodesic Witten diagram prescription of \cite{Kraus1}. The key difference when we are computing semi-classical Virasoro blocks \cite{Kraus} is that the propagators and geodesics of the light and exchanged operators are now computed in the backreacted geometry created by the heavy operators. 

We want to calculate this in the heavy-light limit wherein $h_L, h, h_H/c$ are fixed as $c\rightarrow\infty$. We will consider the case where the two light operators have the same scaling dimension $h_L$, and so do the two heavy operators, $h_H$. It is straightforward to generalize the prescription to incorporate the case when they are different, but this is not needed to capture the bulk physics that is of interest to us.

\subsection{The Conical Defect Prescription}

In \cite{Kraus}, the above integral was evaluated when the heavy operator is a conical defect. This means that the heavy operator scaling dimension is below the BTZ threshold, $h_H < c/24$. The resulting expression for the semi-classical Virasoro block can then be analytically continued in $h_H$ above the BTZ threshold as well. This is perfectly sufficient if one's goal is only to obtain the semi-classical block, but it has the drawback that it is silent about the bulk mechanics of information loss. We would therefore like to obtain the answer in a form that is more directly reliant on the BTZ geometry. But before that, we discuss some aspects of the calculation of \cite{Kraus}.

The conical defect metric used in \cite{Kraus} is taken in the form
\begin{equation}
    ds^2 = \frac{\alpha^2}{\cos^2{\rho}}\left(\frac{d\rho^2}{\alpha^2} + d\tau^2 + \sin^2{\rho}\;d\phi^2\right) \label{ScaledAdS}
\end{equation}
We note that $\alpha = 1$ corresponds to the global (Euclidean) AdS$_3$ metric. To put the metric in a form adapted to the study of the information paradox, let us first do a coordinate transformation 
\begin{equation}
    r = \alpha\tan{\rho},
\end{equation}
which leads to 
\begin{equation}
    ds^2 = \frac{dr^2}{r^2 + \alpha^2} + (r^2 + \alpha^2)d\tau^2 + r^2d\phi^2.
\end{equation}
This form is perhaps more familiar. It is a conical defect metric, if one takes $\alpha = 1/N$ for $N\in\mathbb{N}$. Furthermore, we also see that a naive analytic continuation of $\alpha$ reproduces the Euclidean BTZ black hole metric -- defining $\alpha^2 = -r_+^2$ gives us 
\begin{equation}
    ds^2 = (r^2 - r_+^2)d\tau^2 + \frac{dr^2}{r^2 - r_+^2} + r^2d\phi^2 \label{EBTZ}
\end{equation}
We will mostly be working with this metric in this paper. Note that when doing Witten diagram calculations for Virasoro blocks, one views these geometries simply as coordinate patches -- $\phi$ and $\tau$ span whole real lines, there is no quotienting. We will discuss this further later, but it ultimately related to the branch cut structure of semi-classical Virasoro blocks and the lack of single-valued-ness\cite{Kraus}. When we use the phrase ``Euclidean BTZ'' in this paper, usually we will mean this patch (as should be clear from the context).

The geodesic Witten diagram calculation was done in a slick way in \cite{Kraus} by leveraging the analyticity properties of the block in the parameter $\alpha$. We present the salient features of their approach, below.
\begin{enumerate}
    \item Recognize that $\alpha = 1$ is the global AdS$_3$ limit. We already know the bulk-bulk and bulk-boundary propagators in global AdS$_3$.
    \item Adopt the prescription that propagators in the global limit together with the substitutions $(\tau, \phi) \rightarrow (\alpha\tau, \alpha\phi)$ are the correct propagators in the $\alpha$-scaled metric in \eqref{ScaledAdS}. The latter is to be viewed as the geometry arising from the backreaction of the heavy $h_H$ operator.
    \item The $\alpha$-scaled metric corresponds to a conical defect metric for $0<\alpha<1$ and a BTZ black hole for $\alpha^2<0$. Choose the former for the calculation, as one can always analytically continue in $\alpha$  to get the answer for the case of black holes. In other words, do the geodesic Witten diagram calculation in the ``simpler'' conical defect geometry (and then analytically continue to $\alpha = ir_+$ when needed).
    \item To compare this answer with CFT$_2$ calculations, one can invert the standard radial quantization map given by
    \begin{equation}
        z = e^{iw}\;\;\text{with}\;\;w = \phi + it
    \end{equation}
    to go from the boundary cylinder to the complex plane. The result is the semi-classical Virasoro block.
\end{enumerate}

Now, while this is a valid way of doing the calculation, one would like to set up a similar calculation directly in the BTZ geometry. Part of the motivation behind this was outlined in the Introduction -- we wish to have a more bulk-based understanding of information loss and the role of the horizon (if any). This is not straightforward in the analytic continuation approach, because even though it is valid for the final result, what it means for intermediate steps is unclear. Analytically continuing the parameters causes geodesics to become complex-valued in the BTZ geometry, for example. A second problem is related to the ``source'' of the geometry: the natural source of the geometry is a heavy operator placed at $\rho = 0$ for the conical defect, but when naively translated for the BTZ case, it implies that the source is at $r = 0$. The interpretation of this Euclidean ``interior"  region is murky at best and it is unclear how one might interpret it geometrically. Fundamentally, the trouble is that because it involves an analytic continuation of a {\em parameter}, obtaining the bulk picture above the BTZ threshold from the bulk picture below the threshold, does not seem straightforward.

To make sense of these questions, we will instead look at a modified prescription for understanding the geodesic Witten diagram. We will find that there is a natural way to do this directly in BTZ, which leads to the correct analytically continued semi-classical Virasoro block. The advantage of this approach is that the bulk geometry is sensible while being above the BTZ threshold.

\section{A BTZ Prescription}

The general philosophy of \cite{Kraus} in computing semi-classical Virasoro blocks is that one lets the heavy operator backreact, and then imagines that the light operator is a geodesic on this backreacted geometry. For this strategy to work, one has to have a convenient central location for the ``heavy" geodesic around which one can imagine that the backreacted geometry exists. In the conical defect calculation, the $\rho=0$ location was such a convenient location for the ``heavy" geodesic. At that location, from the form of the metric in \eqref{ScaledAdS}, it is clear that there is a unique ``heavy" geodesic with the natural affine parameter  $ds^2= \alpha^2 d\tau^2$. A second conceptual ingredient was that the bulk-bulk and bulk-boundary propagators were obtained from the global AdS$_3$ quantities, via the coordinate transformation relating the two spacetimes (a simple rescaling of $\tau$ and $\phi$ in the conical defect case). Note that all these geometries are locally AdS$_3$, so they are all related by coordinate transformations. But it was crucial for the success of the calculations in \cite{Kraus} that these propagators were {\em not} genuine bulk-bulk/boundary propagators in the conical defect -- they were simply coordinate transforms of the AdS$_3$ propagators. This leads to the correct branch cut structure of the semi-classical Virasoro block. The rest of the calculation follows more or less naturally once we make these choices. 

The trouble in doing a similar calculation above the BTZ threshold can now be understood as the observation that $\rho=0$ no longer has a clear role as the central ``heavy" geodesic\footnote{In terms of the Witten diagram components, we will take the $\gamma_{12}$ geodesic connecting the two heavy operators as the ``heavy" geodesic, and the $\gamma_{34}$ geodesic connecting the two light operators will be the light geodesic.}.  A key observation is that this problem can be bypassed by working with the metric \eqref{EBTZ}, if one takes the ``heavy" geodesic to be at the location $r=r_+$.  In the next subsection, we will develop this idea further and show that there is also a natural prescription for the bulk-bulk and bulk-boundary propagators in this language. Together with the geodesics in the  backreacted metric \eqref{EBTZ}, we will then be able to complete the calculation using the formula \eqref{GWD}. The interpretation of the boundary as $r\rightarrow\infty$ regime is still intact, so our light geodesic $\gamma_{34}$ will be anchored there.  The exchanged operator will be a field of conformal dimension $(h,\bar{h})$ between the two geodesics -- more precisely, it will be implemented via a bulk-bulk propagator connecting a point on the light operator geodesic to a point on the heavy operator geodesic.

The final result upon doing the integrations in \eqref{GWD} will be the semi-classical Virasoro block.

\begin{figure}[h!]
    \centering
    \includegraphics[width=\linewidth]{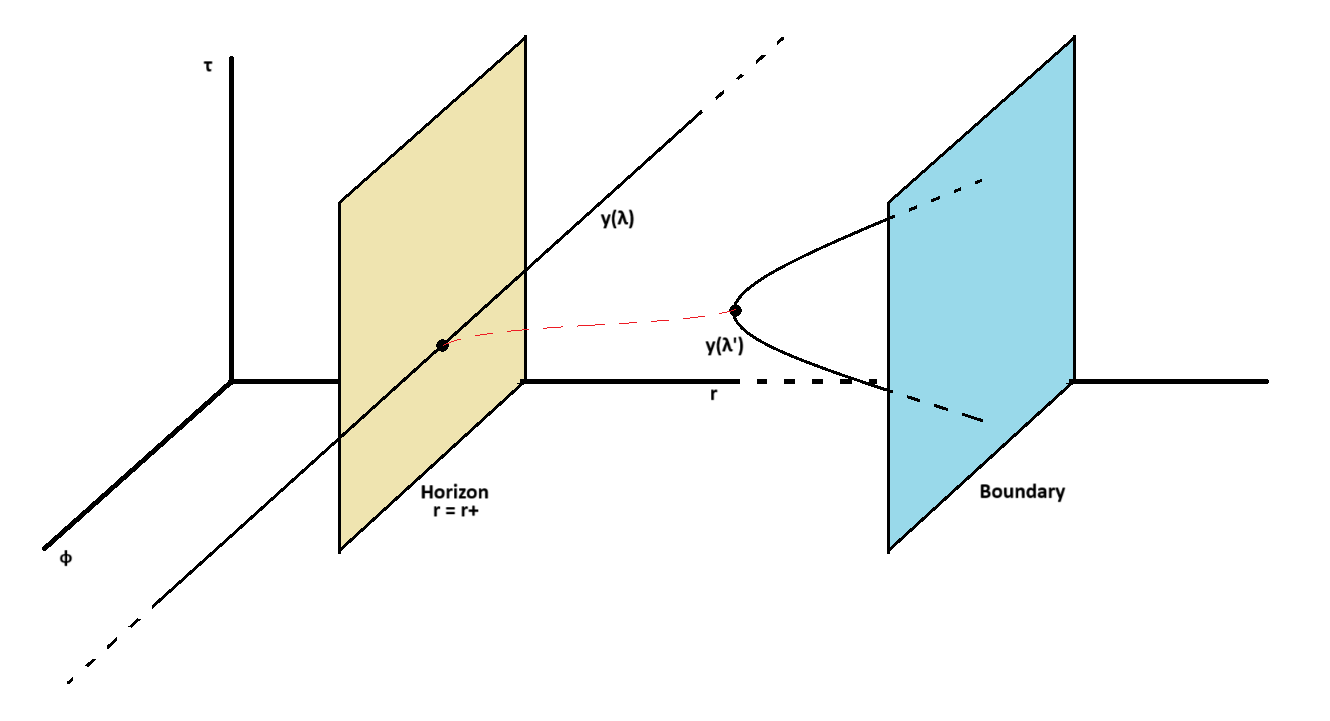}
    \caption[Modified bulk prescription for geodesic Witten diagrams in the semi-classical heavy-light limit]{The BTZ prescription for the bulk geodesic Witten diagram. Even though it is hard to represent in the figure, the metric in the $\tau$-direction has collapsed at $r=r_+$. The endpoints of the $\lambda'$-geodesic are separated in $\tau$ while the $\lambda$-geodesic lies along $\phi$.}
    \label{fig8b}
\end{figure}

\subsection{Set Up}

We start with the metric \eqref{EBTZ} in BTZ coordinates. A bulk point will be denoted by $y = \{r,\tau, \phi\}$, while a boundary point is given by $x = \{\tau, \phi\}$. In both cases, indices or primes will be used, if there are multiple points. As we mentioned already, a key feature is that even though we are working with this new form of the metric, this is really just a patch of AdS$_3$ -- the usual BTZ quotienting is {\em not} imposed. In effect this means that the span of the coordinates $\tau, \phi$ is $(-\infty, +\infty)$. In fact, in the next subsection we will go further, and work with AdS$_3$ propagators in these coordinates (without any image sum) as the appropriate bulk-bulk/boundary propagators. This might seem like a dramatic step, because the coordinate transformation connecting BTZ coordinates and AdS$_3$ is fairly complicated \cite{Ichinose}. But it is important to note that in effect, this was what was done in \cite{Kraus} as well. It was just that in order to reach the coordinates \eqref{ScaledAdS} from global AdS$_3$, one only has to do a simple rescaling, therefore this step may have {\em seemed} more innocuous. 

We place the heavy operator at $r=r_+$ as the source of the geometry. This means that 
\begin{equation}
    \begin{aligned}
        x_1 &= (\tau = \tau_0, \phi = -\infty)\\
        x_2 &= (\tau = \tau_0, \phi = +\infty)\\
    \end{aligned}
\end{equation}
 The value of $\tau_0$ affects none of the results below. This is natural because when one sets $r$ to some constant value, and if that value is $r = r_+$, we have $ds^2 = r_+^2d\phi^2$. So the $\phi$ coordinate has a natural interpretation as an affine parameter at $r = r_+$. Note that this is again a seemingly big difference. The analogous role was taken by $\tau$ in the conical defect calculation. It is crucial to remember here that since we are working in Euclidean signature, there is nothing that picks out ``time'' $\tau$ as the natural affine parameter, since $\phi$ is also non-compact.

Next, we will put the light operators on the boundary $(r' \rightarrow \infty)$ at
\begin{equation}
    \begin{aligned}
        x_3 &= (\tau = \tau_1, \phi = \phi_1)\\
        x_4 &= (\tau = \tau_2, \phi = \phi_2)
    \end{aligned}
\end{equation}
To simplify things, we take $\phi_1 = \phi_2 \equiv \phi=$ constant. This should be compared to the choice made in \cite{Kraus} to keep the boundary geodesic to be on a fixed temporal slice. Because of the (anti-)holomorphic way in which the boundary coordinates enter the expressions, we will be able to re-instate $\phi$ if we choose to. To understand information loss however we are interested in temporal separation. The virtue of our approach is that it naturally leads to boundary geodesics which capture that.

With this understanding, the geometry of the prescription can be schematically presented as in Fig.~\ref{fig8b}. The ``heavy" geodesic is centered on $\phi$ while the light boundary anchored geodesic captures boundary time separation.

\subsection{Propagators}

As we discussed in the previous subsection, the bulk-bulk propagator that we are after is obtained by simply coordinate-transforming the usual AdS$_3$ bulk-bulk propagator. In particular, this means that we do not impose any image sum to incorporate the quotienting. Our bulk-bulk propagator is therefore simply \cite{Ichinose}
\begin{equation}
    G_{bb}(y, y'; 2h) = G^E_F(\xi) = \xi^{2h}\;_2F_1(h, h+\frac{1}{2}; 2h; \xi^2)
\end{equation}
where the $\xi$ is a known simple function of the geodesic distance \cite{Kraus} between the bulk points $y$ and $y'$. The crucial point is that we will view it as a function of the BTZ coordinates \eqref{EBTZ}: 
\begin{equation}
    \xi = \frac{r_+^2}{rr'\cosh{(r_+\Delta\phi)} - \sqrt{r^2 - r_+^2}\sqrt{r'^2 - r_+^2}\cos{(r_+\Delta\tau)}}.
\end{equation}
where $\Delta \phi \equiv \phi - \phi'$ and $\Delta \tau\equiv \tau-\tau'$. To get the bulk-boundary propagator we take the extrapolate limit by sending $r'$ to the boundary $G_{b\partial}(x', y) \equiv \lim_{r'\to\infty} (r')^{2h}G_{bb}(y, y'; 2h)$. The explicit forms of the bulk-boundary/bulk propagators are then
\begin{equation}
    \begin{aligned}
        &G_{b\partial}(x', y) = \left(\frac{r_+^2}{r\cosh{(r_+\Delta\phi)} - \sqrt{r^2 - r^2_+}\cos{(r_+\Delta\tau)}}\right)^{2h}\\
        &G_{bb}(y, y'; 2h) =  \left(\frac{r_+^2}{rr'\cosh{(r_+\Delta\phi)} - \sqrt{r^2 - r_+^2}\sqrt{r'^2 - r_+^2}\cos{(r_+\Delta\tau)}}\right)^{2h} \times \label{bulk-bulk}\\
        &\;\;\;\;\;\times _2F_1\left(h, h+\frac{1}{2}; 2h; \left[\frac{r_+^2}{rr'\cosh{(r_+\Delta\phi)} - \sqrt{r^2 - r_+^2}\sqrt{r'^2 - r_+^2}\cos{(r_+\Delta\tau)}}\right]^2\right)
    \end{aligned}
\end{equation}

\subsection{``Heavy" Geodesic Propagators}

Consider the first piece of the double integral in \eqref{3a}, the one over the parameter $\lambda$. The heavy geodesic sits at $r = r_+$ and extends from $\phi = -\infty$ to $\phi = \infty$. We have the affine parameter
\begin{equation}
    \lambda = r_+\phi.
\end{equation}
The product of the boundary correlators becomes,
\begin{equation}
    \begin{aligned}
        G_{b\partial}(x_1, y(\lambda))G_{b\partial}(x_2, y(\lambda)) &= G_{b\partial}(\phi = -\infty, \phi(\lambda))G_{b\partial}(\phi = \infty, \phi(\lambda))\\
        &= \left(\frac{r_+}{\cosh{(\lambda\rightarrow+\infty)}}\right)^{2h_H}\left(\frac{r_+}{\cosh{(\lambda\rightarrow-\infty)}}\right)^{2h_H}\\
        &\sim 1,
    \end{aligned}
\end{equation}
and thus this piece contributes unity (up to some constant prefactors that we do not worry about).

\subsection{Light Geodesics and Light Geodesic Propagators}

We will first compute the geodesics and then the bulk-boundary propagators (of the light operators) that end on them. Since these geodesic will play a crucial role in our discussion of information loss, we present them in some detail.

\subsubsection{Geodesics}

Since we are working with a constant $\phi$-slice for the light geodesics, we will have two equations to solve to determine the geodesics. They are
\begin{equation}\label{3b}
    (r^2 - r_+^2)\dot{\tau}^2 + \frac{1}{(r^2 - r_+^2)}\dot{r}^2 = 1,
\end{equation}
and
\begin{equation}\label{3c}
   \ddot{\tau} + \left(\frac{2r}{r^2 - r_+^2}\right)\dot{r}\dot{\tau} = 0,
\end{equation}
where $\dot{z}$ signifies the derivative of ${z}$ with respect to $\lambda'$.

Integrating the last equation as
\begin{equation}\label{3d}
    \dot{\tau} = \frac{c_1}{r^2 - r_+^2}
\end{equation}
(where $c_1$ is some constant of integration) and using it in \eqref{3b} to solve the latter, we get
(after suitable redefinitions of the affine parameter, $\lambda' \rightarrow a \lambda' +b$), 
\begin{equation}\label{3e}
   r = \sqrt{r_+^2 + c_1^2}\cosh{\lambda'}.
\end{equation}
We are choosing $\lambda'\rightarrow-\infty$ to correspond to $x_3$ and $\lambda'\rightarrow\infty$ to correspond to $x_4$. Note that $\lambda'\rightarrow \pm\infty$ gives us $r\rightarrow\infty$ as desired. Using \eqref{3e} in \eqref{3d}, we also get 
\begin{equation}\label{3f}
    \tau + c_2 = \frac{1}{r_+}\tan^{-1}{\left(\frac{r_+}{c_1}\tanh{\lambda'}\right)}
\end{equation}

Imposing the boundary conditions for $\tau$ at $\lambda' = \pm\infty$ as $\tau_1, \tau_2$, we can fix $c_1, c_2$:  
\begin{equation}
    \begin{aligned}
        c_1 &= r_+\cot{\left[\frac{r_+}{2}(\tau_2 - \tau_1)\right]}\\
        c_2 &= -\frac{(\tau_1 + \tau_2)}{2}.\\
    \end{aligned}
\end{equation}
The final solution for the geodesic is then
\begin{equation}\label{3g}
    \begin{aligned}
        r(\lambda') &= r_+\csc{\left[\frac{r_+}{2}(\tau_2 - \tau_1)\right]}\cosh{\lambda'}\\
        \tau(\lambda') &= \frac{(\tau_1 + \tau_2)}{2} + \frac{1}{r_+}\tan^{-1}{\left(\tan{\left[\frac{r_+}{2}(\tau_2 - \tau_1)\right]}\tanh{\lambda'}\right)}\\
        \phi(\lambda') &= \textrm{const.}
    \end{aligned}
\end{equation}

It is convenient to introduce the following variables:
\begin{equation}\label{3j}
    \begin{aligned}
        \theta(\lambda') &= r_+\tau(\lambda') - r_+\left(\frac{\tau_1 + \tau_2}{2}\right)\\
        \psi &= \frac{r_+}{2}\left(\tau_2 - \tau_1\right) \equiv \frac{r_+\tau}{2}
    \end{aligned}
\end{equation}
The geodesics can be written in a condensed form after these re-definitions:
\begin{equation}
   \begin{aligned}
        r= r_+\csc{\psi}\cosh{\lambda'}, \ \ 
        \tan{\theta} = \tan{\psi}\tanh{\lambda'} \label{psi-theta}
    \end{aligned}
\end{equation}

A key observation from the equations is that these geodesics with time-separated endpoints only exist when $0 \le \tau \le \frac{\pi}{r_+}$. At the upper limiting value, they effectively fall into the horizon. This timescale is half the Euclidean periodicity of the usual BTZ black hole cigar.  Our computation of the semi-classical Virasoro blocks will be done in the $0 \le \tau \le \frac{\pi}{r_+}$ regime, but once we have the result, we will be able to extend it beyond that limit by analytic continuation. These observations will have important consequences for information loss, which we will discuss later.


\subsubsection{Propagators}

The light geodesic propagators take the form 
\begin{equation}
    \begin{aligned}
        G_{b\partial}(\tau_1, y(\lambda')) &= \left(\frac{r_+^2}{r(\lambda') - \sqrt{(r^2(\lambda') - r^2_+)}\cos{[r_+(\tau(\lambda') - \tau_1)]}}\right)^{2h_L}\\
        &= \left(\frac{r_+}{\csc{\psi}\cosh{\lambda'} - \sqrt{(\csc^2{\psi}\cosh^2{\lambda'} - 1)}(\cos(\theta +\psi)}\right)^{2h_L}
    \end{aligned}
\end{equation}
which using our definition of $\theta$ and various hyperbolic/trigonometric identities can be brought to the simple form
\begin{equation}
    G_{b\partial}(\tau_1, y(\lambda')) = \left(r_+\frac{e^{-\lambda'}}{\sin{\psi}}\right)^{2h_L}
\end{equation}
We can do the same for the other boundary point and we observe that since we have $\cos{(\theta - \psi)}$, we get $e^{+\lambda'}$ in the numerator:
\begin{equation}
   G_{b\partial}(\tau_2, y(\lambda')) = \left(r_+\frac{e^{+\lambda'}}{\sin{\psi}}\right)^{2h_L}.
\end{equation}
Putting these together, we have 
\begin{equation}
    \begin{aligned}
        G_{b\partial}(x_3, y(\lambda'))G_{b\partial}(x_4, y(\lambda')) &= G_{b\partial}(\tau_1, y(\lambda'))G_{b\partial}(\tau_2, y(\lambda'))= \left(\frac{1}{\sin^2{\psi}}\right)^{2h_L}.
    \end{aligned}
\end{equation}
Again, this is up to some constant prefactors.

\subsection{Bulk-Bulk Propagator}

The last piece of the geodesic Witten diagram is the bulk-bulk propagator \eqref{bulk-bulk}, which is the field exchanged between the light and heavy operator geodesics. The primed coordinates belong to the light operator geodesic, while the unprimed ones are on the heavy operator geodesic. We can make use of special properties of the individual geodesics to simplify the expression considerably.

We note that since $r = r_+$ for the heavy geodesic, the factor in front of the $\cos{(r_+\Delta\tau)}$ vanishes, and so, this function is only dependent on $r, r', \phi, \phi'$. We can simplify things even more if we take $\phi' = 0$. As it stands, the light operator geodesic is on a constant $\phi'$ surface, so choosing a constant value amounts to fixing that surface. Furthermore, since $\lambda = r_+\phi$ for the heavy geodesic, we can think of this constant $\phi'$ of the light operator geodesic as a constant added to the affine parameter $\lambda$ of the heavy geodesic. In any case, what remains inside the hyperbolic cosine is simply $\lambda$. Using this, we get 
\begin{equation}
        G_{bb}(y, y'; 2h) =  \left(\frac{\sin{\psi}}{\cosh{\lambda'}\cosh{\lambda}}\right)^{2h}\;_2F_1\left(h, h+\frac{1}{2}; 2h; \left[\frac{\sin{\psi}}{\cosh{\lambda'}\cosh{\lambda}}\right]^2\right),
\end{equation}
where we have used \eqref{psi-theta} to write $r'(\lambda')$ as $\frac{r_+ \cosh \lambda'}{\sin \psi}$.

\subsection{Assembling the Geodesic Witten diagram}

With all the pieces put together, we can now calculate the geodesic Witten diagram for the semi-classical Virasoro block:
\begin{equation}
    \begin{aligned}
        \mathcal{W}_{2h,0}(\tau, \phi)
        &\sim \int_{-\infty}^{+\infty} d\lambda\int_{-\infty}^{+\infty} d\lambda'\;\left(\frac{1}{\sin^2{\psi}}\right)^{2h_L}\left(\frac{\sin{\psi}}{\cosh{\lambda'}\cosh{\lambda}}\right)^{2h} \times\\
        &\;\;\;\;\;\;\;\;\;\;\;\;\;\;\;\;\;\;\;\;\;\;\;\;\;\;\;\;\;\;\;\;\;\;\;\times\;_2F_1\left(h, h+\frac{1}{2}; 2h; \left[\frac{\sin{\psi}}{\cosh{\lambda'}\cosh{\lambda}}\right]^2\right)
    \end{aligned}
\end{equation}
Our choice of prescription has led to the $\phi$ dependence to drop out, and since we are interested in the $\tau$ dependence of the Virasoro blocks anyway, we will denote this as $\mathcal{W}_{2h,0}(\tau)$ to make the time dependence explicit.

Because of the variable re-definitions we have made in \eqref{psi-theta}, the above integrand has turned essentially into the same form found in \cite{Kraus}. This allows us to use their technology to do the integral, and the end result is (after replacing $\psi$ according to \eqref{3j})  
\begin{equation}
    \mathcal{W}_{2h,0}(\tau) \sim \left[\sin{\left(\frac{r_+\tau}{2}\right)}\right]^{2h - 4h_L}\;_2F_1\left(h, h; 2h; 1 - e^{ir_+\tau}\right)\cdot\;_2F_1\left(h, h; 2h; 1 - e^{-ir_+\tau}\right),
\end{equation}
This is an expression for the Virasoro block as desired. An interesting feature of this result is that because our calculation  works with a geodesic in the time direction, the result is naturally holomorphically factorized with complex conjugate pieces. The calculation in \cite{Kraus} used a geodesic at a constant time that was separated in $\phi$, and therefore the holomorphic factorization had a distinct origin.

As in \cite{Kraus}, it is also possible to obtain this result by a single integral. We write the geodesic Witten diagram  as
\begin{equation}
    \begin{aligned}
        \mathcal{W}_{2h,0}(\tau, \phi)
        &= \int d\lambda'\;G_{b\partial}(x_3, y(\lambda'))G_{b\partial}(x_4, y(\lambda'))\Psi(y(\lambda'))
    \end{aligned}
\end{equation}
where $\Psi(y')$ is given by the inner integral
\begin{equation}
    \begin{aligned}
        \Psi(y') &= \int d\lambda\;G_{b\partial}(x_1, y(\lambda))G_{b\partial}(x_2, y(\lambda))\times G_{bb}(y(\lambda), y'; 2h)\\
        &= \int d\lambda\;G_{bb}(y(\lambda), y'; 2h)
        = r_+\int d\phi\;G_{bb}(y(r_+\phi), y'; 2h)
    \end{aligned}
\end{equation}
The latter expression implies that this $\Psi$ solves the bulk wave equation (note that for us, it is in BTZ coordinates): $\Box' \Psi = 4h(h-1) \Psi$. The normalizable solution to this is well-known:
\begin{equation}
    \Psi(y') = \left(\frac{r_+^2}{r'^2}\right)^h\;_2F_1\left(h, h , 2h, \frac{r_+^2}{r'^2}\right)
\end{equation}
Pulling this back onto the light geodesic means letting $y'\equiv y'(\lambda')$. Substituting \eqref{3g} and plugging it back, we have
\begin{equation}
    \mathcal{W}_{2h,0}(\tau, \phi) = \int d\lambda'\;\left(\frac{1}{\sin^2{\psi}}\right)^{2h_L}\left(\frac{\sin^2{\psi}}{\cosh^2{\lambda'}}\right)^h\;_2F_1\left(h, h , 2h, \frac{\sin^2{\psi}}{\cosh^2{\lambda'}}\right)
\end{equation}
This is again a form that has appeared in \cite{Kraus} and leads to the correct semi-classical Virasoro block.

\section{Comments on Virasoro Blocks}\label{VirBloDisc}

We will summarize some observations here about semi-classical and finite-$c$ Virasoro blocks. These observations seem significant for an eventual understanding of the information paradox and are important for the philosophy of this paper. Some of the comments will also be relevant for the discussions in Section \ref{Disc} and Appendix \ref{AppFourier}.

Finite-$c$ Virasoro blocks are difficult to study analytically. They were studied numerically in Euclidean and Lorentzian signatures in \cite{KapN}. In Lorentzian signature it was found that the block, which has an exponential decay at early times, switches to a power law decay $\sim t^{-3/2}$ at about the Page time   $\sim \mathcal{O}(S)$. This switching over from exponential to $\sim t^{-3/2}$ decay at late (but not exponentially late) times, is a standard feature of strongly coupled quantum chaotic systems and random matrices, and is familiar from studies of the Spectral Form Factor (SFF) \cite{Cotler}. The result of \cite{KapN} is remarkable because it makes it very plausible that the exponential decay in the full correlator also gets arrested in some manner at the Page time. This is an honest-to-God ``black hole microstate" result\footnote{All other resolutions that we know of, are in {\em models} for black hole microstates on the boundary \cite{Cotler} or in the bulk \cite{tHooft, Vaibhav1, Vaibhav2}.}. It makes it overwhelmingly likely that there exist mechanisms in AdS$_3$/CFT$_2$ that kick in at the Page time to resolve Maldacena's information paradox \cite{Maldacena}. This curbing of thermal decay at the Page time has implications. It shows (for example) that applying the thermal correlator for  interior reconstruction beyond the Page time, may not be meaningful \cite{Vaibhav2}.

It is also suggested in \cite{KapN} that the $\sim t^{-3/2}$ behavior persists for arbitrarily long times. It may be worth making doubly sure that this last claim (which is again numerical) is fully reliable, because a significant amount of physics rests on it, as we will now argue. First, there is an interesting and very large timescale $\mathcal{O}(e^S)$ called the dip time, where $\sim t^{-3/2}$ behavior often flips over to a ramp like behavior in strongly quantum chaotic systems. Second, it means that the ramp behavior that is universally expected in these systems is not present in Virasoro blocks, implying that ramp physics is controlled by the 3-point function coefficients in the HLLH correlator. It will be very interesting to understand a mechanism by which this can happen. Third, it was noted in \cite{KapN} that a $\sim t^{-3/2}$ behavior is suggestive of Wigner semi-circle and random matrices. It would be interesting to understand what causes this behavior to appear, but not the ramp behavior (which is also a standard feature of random matrices). It has been noted in \cite{Riemann1, Riemann2} that the power law decay of the dip need not always be $\sim t^{-3/2}$, even if the linear ramp is present. The case of the numerical Virasoro block seems to indicate that the implication need not hold the other way either. 

The $\sim t^{-3/2}$ behavior expected in the correlator and explicitly seen to be present in finite-$c$ Virasoro blocks indicates that in momentum space, the correlator has a branch cut behavior $|\omega|^{1/2}$ for small $\omega$'s. Depending on the the true post-$\mathcal{O}(e^S)$ time behavior, there may be refinements to this observation. A striking and well-known feature of semi-classical Virasoro blocks is that they have a simple connection to the image sum form of BTZ Green functions -- if we only retain the $n=0$ contribution in the mirror sum, the object that one gets is essentially the semi-classical Virasoro block. This is discussed in \cite{KapInfo, KapProp}. Going back and forth between position and momentum space when there are multiple time scales is often difficult. In fact, even for the semi-classical 2-point function (where the only scale in the problem is the Hawking temperature) explicitly doing the Fourier transform is non-trivial, and is generally not addressed in the literature (but see \cite{CK} for some closely related comments). The bulk version of this calculation brings out some noteworthy features, so we do it explicitly in an Appendix. It may be interesting to generalize this calculation to the case when none of the image contributions are retained and we are working with a (bulk version of the) semi-classical Virasoro block.

\begin{figure}[h!]
  \centering
  \includegraphics[width=\textwidth]{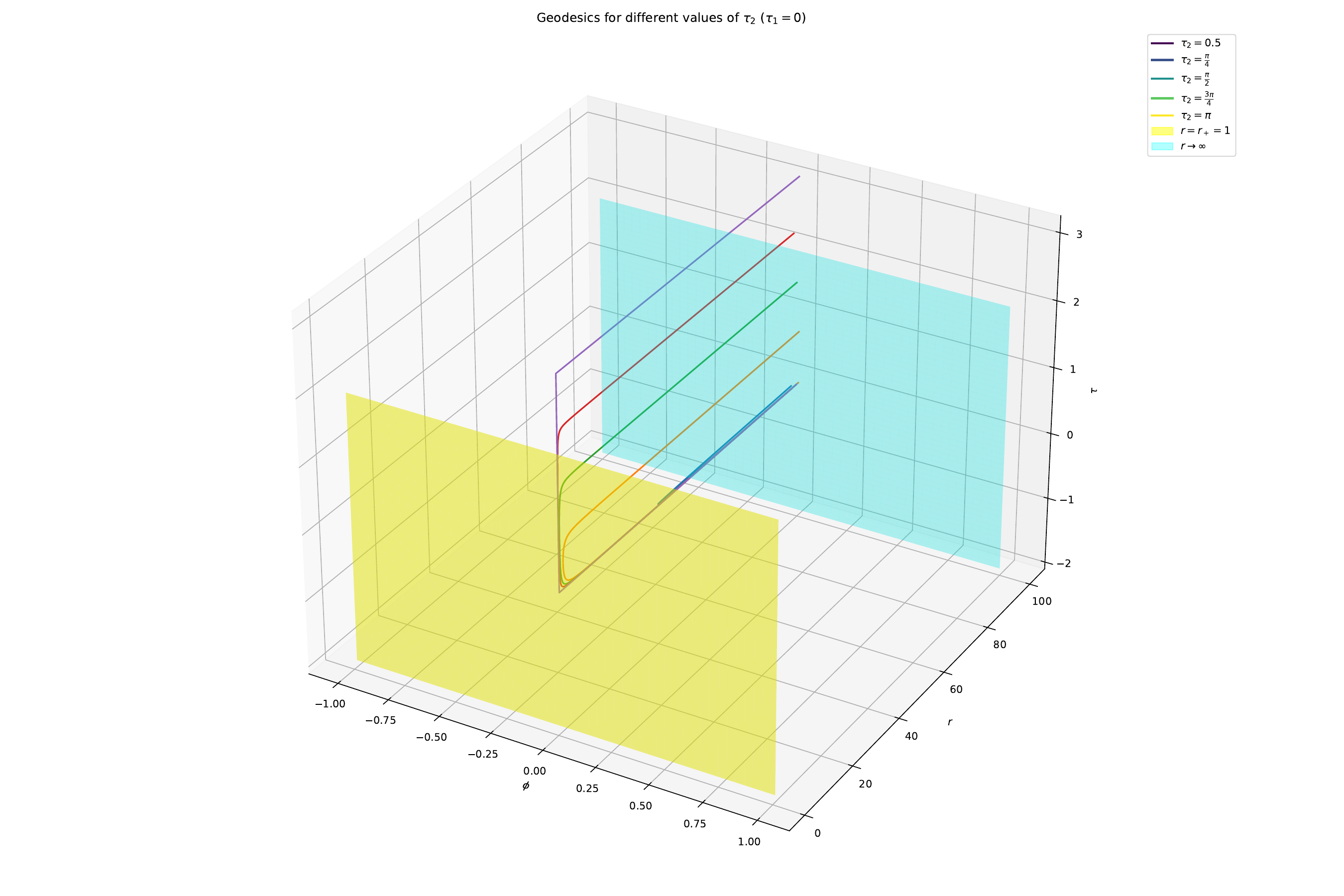}
  \caption[Geodesics in the new bulk prescription I]{Geodesic plots for different $\tau = \tau_2-\tau_1$ values. We have set $r_+=1$. The boundary is given in blue, while the horizon is given in yellow. Observe how as $\tau\rightarrow\pi$, the geodesics go closer to the horizon and also get a more rectangular profile. At $\tau = \pi$ they hit the horizon.}
  \label{fig4b}
\end{figure}

\begin{figure}[h!]
  \centering
  {\includegraphics[width=0.8\textwidth]{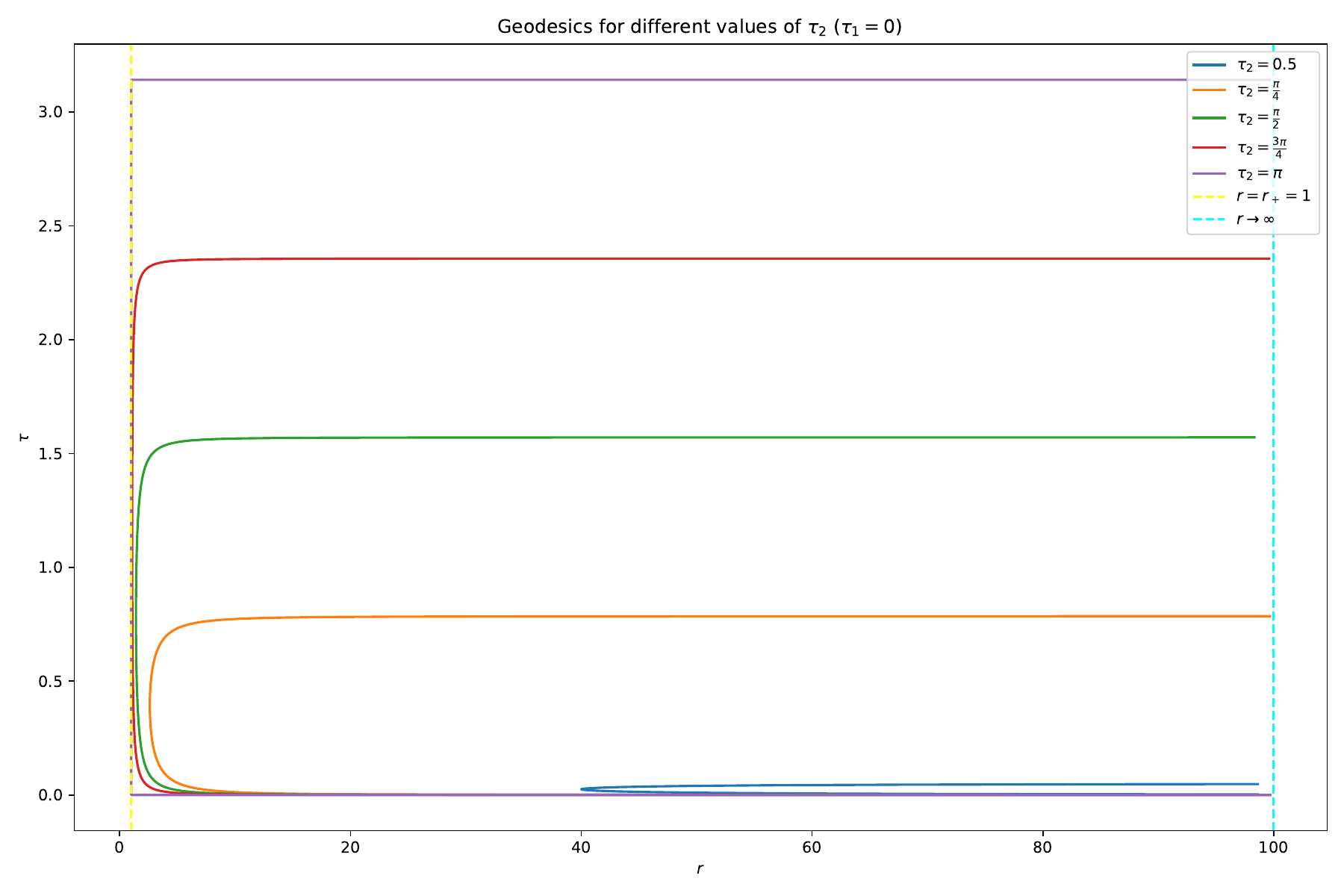}}
  \caption[Geodesics in the new bulk prescription II]{Geodesic plots for different $\tau$ values. The boundary is given in blue, while the horizon is given in yellow. Observe how as $\tau\rightarrow\pi$, the geodesics go closer to the horizon and also get a more rectangular profile.}
  \label{fig4e}
\end{figure}

Much of the above discussion on Virasoro blocks is in Lorentzian signature, while our focus in this paper is fundamentally Euclidean. In the semi-classical limit, the exponential Lorentzian decay and the Euclidean periodic singularities are related by Wick rotation in a simple way -- one simply replaces the Lorentzian $t \rightarrow -i \tau +i\epsilon$ where $\tau$ is Euclidean. The situation is likely to be significantly more complicated at finite-$c$. At finite-$c$ the numerical approach of \cite{KapN} showed that semi-classical Virasoro blocks break down completely in a finite region around the forbidden Euclidean singularities. While this makes the full understanding of the resolution a tall order at this time, we can still hope to make some progress by considering the origin of Euclidean information loss {\em from the bulk,} via semi-classical blocks. This is our modest goal in this paper. In the following section, we will provide a bulk understanding of some basic qualitative features of the numerical Euclidean blocks in figure 15 of \cite{KapN}.

\section{Information Loss from Geodesics at the Euclidean Horizon}\label{Disc}

Before we go into details, we urge the reader to have a look at Figure 15 of \cite{KapN}, which contains plots of semi-classical and finite-$c$ Euclidean blocks (the latter computed numerically for a few values of $c \gg 1$) against Euclidean time\footnote{The plots there are against the uniformizing variable $q$, and so the periodicity is not manifest, but this is not important for us. The location of the half-period and period are clear from the plots.}. It is clear that semi-classical blocks have periodic singularities, while the exact blocks do not have any singularities other than the coincidence singularity at the origin. We will also note that even for relatively low values of $c$ (=30) the deviation between exact and semi-classical blocks is unnoticeably small before the half-period, while it begins to be noticeable after that. We will see that the geodesic Witten diagram gives an understanding of this feature.

It is instructive to plot the geodesics contributing to the semi-classical Virasoro blocks in our calculation, as a function of the temporal separation of its initial and final points. This is provided in Figs.~\ref{fig4b} and \ref{fig4e}. Note that $\beta = 2 \pi/r_+$. If we look at what happens around $\tau = \pi/r_+$ for our geodesics (which is half the period), we observe that this is the time at which the geodesic straddles the horizon (refer to Figs.~\ref{fig4b} and \ref{fig4e}).

These plots illustrate a precise way in which the horizon can be held responsible for Euclidean information loss -- information loss starts becoming substantial when the light operator geodesic starts probing the horizon radius associated to the heavy operator. It is also interesting to note that the calculation is not merely loosely sensitive to the periodicity timescale $\beta$, but is tied specifically to the the half-period due to the geometry of the geodesic. The semi-classical Virasoro block is naturally viewed as being computed by our geodesic Witten diagram in this region, and then extended to longer times by analytic continuation. These observations can be taken as suggestive that we need modifications to the bulk and/or the probe geodesic prescription, starting around the horizon scale, if we wish to avoid information loss. Some discussions of this kind, will be reported elsewhere.

\section{Acknowledgments}  

We thank Ujjwal Basumatary, ChatGPT, Diptarka Das, Per Kraus, Pradipta Pathak and Shivrat Sachdeva for discussions.

\appendix

\section{4-Point Functions and Virasoro Blocks}\label{AppVirasoro}

Any 4-point correlator in a 2D CFT can be written as a sum over Virasoro blocks by fusing operators pairwise in an OPE. We provide a fairly self-contained introduction to this fact, in the concrete setting of the HLLH 4-point function in this Appendix. 

Explicitly, the correlator is 
\bea
\langle O_H O_L O_L O_H \rangle \equiv \langle  O_H (\infty, \infty) O_L(1,1) O_L(z, \bar z) O_H(0,0) \rangle.
\eea
It is often conventional to suppress the anti-holomorphic coordinates. Three of the four points above were fixed using global conformal invariance (M\"obius transformations). We can take the points to be radially ordered starting from the fourth operator (the one at the origin), the third (the one at $z$),  the second (the one at 1) and then the first operator (the one at infinity). This means that we are taking
 $0<|z|<1$ here\footnote{This might seem like loss of generality. But it is worth noting that in the $s$-channel (to be described below) we will be doing OPEs between the third and fourth operators. The OPE convergence condition around the fourth operator is precisely $|z|<1$. If we wish, we can switch channels and access other regions of the complex plane. Bootstrapping requires (and it can be checked) that there are regions in the complex plane where both channels are simultaneously convergent.}. The operator at infinity above requires a bit of care:
\bea
 O_H (\infty, \infty) \equiv \lim_{w\rightarrow \infty} w^{2 h_H} {\bar w}^{2{\bar h}_H} O_H(w, \bar w). \label{inftyOp}
 \eea
In other words, we define the operator at infinity (i.e., on the North pole of the sphere) as the operator that maps to the origin under inversions. This is necessary because $\infty$ is not an element of $\IC$, while the other three points are, and so we have to define $\infty$ as a suitable limit on $\IC$. 

We will do the pairwise fusion between first and second operators as well as the third and fourth operators. This is called the $s$-channel. Before going ahead with our main goal of illustrating the emergence of Virasoro blocks, let us note that pairwise fusion can be done in three different ways. This is because there are three ways to pair the first operator (there are three {\em other} operators in a 4-point function), and once this is done, the remaining two operators have a unique pairing. The three channels are called $s, t$ and $u$, and for concreteness, we will use the $s$-channel to illustrate the emergence of blocks. We first write the OPE between 3 and 4:
\begin{equation}
\mathcal{O}_L(z, \bar z)\, \mathcal{O}_H(0,0) 
= \sum_p C_{LHp} \, z^{h_p - h_L - h_H} \bar z^{\bar h_p - \bar h_L - \bar h_H} 
\left[ \mathcal{O}_p(0) + \sum_{n, \bar n > 0} z^n \bar z^{\bar n}\, 
\beta_{n, \bar n}^{(p)}\, \mathcal{O}_p^{(n, \bar n)}(0) \right]. \label{3-4OPE}
\end{equation}
The outermost sum is over primaries, $p$. The key point is that the once we know the OPE coefficients for the primaries, the OPE coefficients for the Virasoro descendants can be determined from the conformal algebra. Apart from the $z$ and $\bar z$ needed to match the extra scaling dimension of the descendants, these coefficients we have denoted by the $\beta_{n, \bar n}^{(p)}$'s above. They are fully determined by the kinematics of the Virasoro algebra and the primary/descendant structure, even though they are too complicated to write down usefully. As a result of the left-right tensor product structure of Virasoro modules, $\beta$ factorizes as $\beta^{(p)}_{n,\bar n} =\beta^{(p)}_{n} \beta^{(p)}_{\bar n} $. Therefore the OPE in \eqref{3-4OPE} can be re-written as 
\begin{equation}
\mathcal{O}_L(z,\bar z)\,\mathcal{O}_H(0)
= \sum_{p} C_{LHp}
  \sum_{n,\bar n \ge 0}
    \beta_{n}^{(p)}\,\beta_{\bar n}^{(p)}
    \;z^{\,h_p + n - h_L - h_H}\;
    \bar z^{\,\bar h_p + \bar n - \bar h_L - \bar h_H}\;
    \mathcal{O}_p^{(n,\bar n)}(0),
\end{equation}
We have adopted the notation that $\mathcal{O}_p(0)=\mathcal{O}_p^{(0,\bar 0)}(0)$. Plugging this into the 4-point function, we are lead to evaluate a sum over 3-point functions involving $\mathcal{O}_p^{(n, \bar{n})}(0)$. From the general structure of CFT 3-point functions, we find\footnote{The \eqref{inftyOp} definition is important here.} 
\begin{equation}
\langle \mathcal{O}_H(\infty) \, \mathcal{O}_L(1) \, \mathcal{O}_p(0) \rangle = C_{HLp}, \ \ \langle \mathcal{O}_H(\infty) \, \mathcal{O}_L(1) \, \mathcal{O}_p^{(n, \bar{n})}(0) \rangle = C_{HLp} \, \gamma_{n}^{(p)}\,\gamma_{\bar n}^{(p)} \label{1-2OPE}
\end{equation}
for primaries and descendants. The $\gamma$'s are the analogues of $\beta$'s for the 1-2 OPE expansion around $\infty$, and are again completely determined in terms of the conformal algebra -- there is no information about 3-point function coefficients in them. Putting it into the four‐point function (which leads to the sum over $p$ on the RHS), gives
\begin{eqnarray}
\langle  O_H (\infty, \infty) O_L(1,1) O_L(z, \bar z) O_H(0,0) \rangle =  \hspace{2.5in} \nonumber\\ \hspace{0.75in} = \sum_{p} C_{LHp}\,C_{HLp}\,z^{-h_L - h_H}\,\bar z^{-\bar h_L - \bar h_H}
\sum_{n,\bar n \ge 0}
  \bigl[\beta_{n}^{(p)}\gamma_{n}^{(p)}\bigr]
  \bigl[\beta_{\bar n}^{(p)}\gamma_{\bar n}^{(p)}\bigr]
  \,z^{\,h_p + n}\,
  \bar z^{\,\bar h_p + \bar n}.
\end{eqnarray}
Now we define Virasoro blocks simply as
\[
\mathcal{V}_{h_p}(z)
= z^{h_p-h_L - h_H}\sum_{n=0}^\infty (\beta_{n}^{(p)}\gamma_{n}^{(p)})\,z^n,
\quad
\bar{\mathcal{V}}_{\bar h_p}(\bar z)
= \bar z^{\bar h_p-\bar h_L - \bar h_H}\sum_{\bar n=0}^\infty (\beta_{\bar n}^{(p)}\gamma_{\bar n}^{(p)})\,\bar z^{\bar n}.
\]
which leads to the  Virasoro block decomposition of the 4-point function:
\[
\bigl\langle \mathcal{O}_H(\infty)\,\mathcal{O}_L(1)\,
\mathcal{O}_L(z,\bar z)\,\mathcal{O}_H(0)\bigr\rangle
= \sum_{p} C_{LHp}\,C_{HLp}\,
  \mathcal{V}_{h_p}(z)\,\bar{\mathcal{V}}_{\bar h_p}(\bar z) = \sum_{h, \bar h} P_{h, \bar h} \,
  \mathcal{V}_{h}(z)\,\bar{\mathcal{V}}_{\bar h}(\bar z).
\]
In the last step we have used the permutation symmetry of 3-point functions to write $C_{LHp}=C_{HLp}$ and defined $P_{h,\bar h} \equiv C^2_{HLp}$. The sum over $h, \bar h$ in the last step is now understood to be over the exchanged primaries, all other dependencies are suppressed. Sometimes it is useful to explicitly split off the vacuum piece ($h=\bar h =0$) in the sum on the RHS as $\mathcal{V}_{0}(z)\,\bar{\mathcal{V}}_{0}(\bar z)$, in which case the remaining sum in $(h, \bar h)$ is over all operators other than the identity. (We can set $P_{0,0} =1$ by choosing the normalization of the 2-point function to be 1.)

It is important to note that the Virasoro block $\mathcal{V}_{h}(z)$ that we have written has dependence on the external operators $h_H, h_L$, the exchanged operator $h$, as well as the central charge $c$. The dependence on the central charge arises from the dependence of $\beta$ and $\gamma$ on $c$ -- the latter is a consequence of the fact that in order to determine the $C_{ijk}$'s for the descendants, we need to use the Virasoro algebra.

Virasoro blocks at finite-$c$ in large-$c$ holographic theories contain an enormous amount of information about quantum gravity. But making useful statements about them, despite their Platonic nature, is hard. One can do better in the semi-classical limit, where explicit forms of the blocks can be determined from global blocks \cite{KapClass, Kraus, KapInfo}. These are what we have relied on, in the present paper.

\section{A Basic Fourier Transform}\label{AppFourier}

One way to investigate the possibility of structure at the horizon is by changing the familiar infalling boundary conditions in some way, and seeing its impact on exterior correlators. It was noticed in \cite{Vaibhav1, Vaibhav2} that exterior correlators can be indistinguishable (up to exponentially suppressed corrections in the entropy) from smooth horizon correlators, even when the boundary conditions are dramatically different --- as long as these differences are at about a Planck length or so from the horizon\footnote{The papers \cite{Vaibhav1, Vaibhav2} studied Dirichlet (brick-wall) boundary conditions, but we suspect the message may apply for larger classes of boundary conditions.  There are also some caveats which are discussed in these papers, but we do not believe these caveats alter the ultimate message. See also related work in \cite{Adepu, Sumit1, Pradipta, Banerjee0, Banerjee}.}. A possible strategy for probing finite-$N$ effects therefore is to try and modify bulk physics at $\sim$ Planck length from the horizon in some inspired way, and see whether one can identify the resulting features in finite-$N$ (or finite-$c$) aspects of CFT correlators or Virasoro blocks.

There are clearly many (non-perturbative) difficulties in adopting such a path. But quite apart from conceptual challenges, there are also some basic pragmatic difficulties.  One such problem is that in the bulk, we typically study correlators in black hole spacetimes by solving wave equations, and going to Fourier space. These correlators, at least for light large-$N$ factorized operators, are easy to compute in momentum space -- they are approximately free fields. This is why it was possible to study brick-wall correlators in \cite{Vaibhav1, Vaibhav2} by simply changing boundary conditions near the horizon. However, the correlators and blocks that one works with in the CFT are often in position space. In order to connect the two one will have to understand general aspects of these Fourier transforms. It turns out that even in the infalling/smooth horizon case, explicit Fourier transforms are challenging.

In this Appendix, we would like to investigate this explicitly for the BTZ black hole with a smooth horizon. There are a few things that make BTZ a nice warm up for our purposes. The first is that the bulk radial wave equation is exactly solvable (unlike in higher dimensions), so one can determine the Fourier space correlator in an explicit form containing Gamma functions and hypergeometric functions \cite{FestucciaThesis}. The second is that (unlike all other black holes?) we can also determine the BTZ correlator directly in position space, by invoking the fact that BTZ is a quotient of AdS$_3$ and using an image sum of the AdS$_3$ correlator \cite{Ichinose}. Both these results are well-known in the literature, but as far as we are aware, an attempt to relate the two by direct Fourier transformation has not been made in publicized work\footnote{The closest result we know is a result in \cite{CK} for the boundary BTZ correlator -- here we will generalize it to the bulk where the full image sum structure is present.}. A third reason for studying BTZ is that finite-$N$/finite-$c$ effects are likely to be most accessible in AdS$_3$/CFT$_2$. In higher dimensions, finite-$N$ effects (especially in the context of heavy states) in the dual CFT are very poorly understood\footnote{In higher dimensions, the bulk side is also harder, because even the radial wave equation is not solvable in a useful way. But we believe this may not be a forbidding problem, because correlators may still be computable because connection formulas for wave equations are tractable due to a relationship with (...wait for it!...) fusion rules for Virasoro blocks \cite{Naidiuk, Bonelli}.}.

In the rest of this Appendix, our goal will simply be to relate the (known) position and momentum space BTZ correlators via Fourier transform. We hope that some of the insights we will learn below by doing this, have broader applicability. It may also be interesting to try and understand this calculation when the image sum is restricted to just the $n=0$ piece -- this is an object closely related to the Virasoro block. 

Even though the various 2-point functions (Wightman, retarded, Feynman, ...) contain equivalent information, it turns out that the Fourier transform is most conveniently accomplished via the so-called ``2-sided correlator'' $G_{12}$ in the black hole background. This object can be related to more familiar Green functions via
\begin{equation}
    G_F(t) = \theta(t)G_+(t) + \theta(-t)G_-(t), \label{Wightman}
\end{equation}
and
\begin{equation}
    G_{12}(t) = G_{+}\left(t - i\frac{\beta}{2}\right),  \label{G12}
\end{equation}
where $G_F$ is the Feynman propagator and $G_{\pm}$ are the Wightman functions. In the eternal black hole geometry, $G_{12}$ is the correlator between points on regions I and III. We will first present this quantity in momentum space, then position space, and then do the Fourier transform of the position space quantity to demonstrate its equivalence to the momentum space expression.

\subsection{Momentum Space}

To discuss momentum space Green functions in the BTZ black hole, we follow the notation of Chapter 2 and Appendix A of ~\cite{FestucciaThesis}. The Lorentzian BTZ metric is
\begin{equation}
    ds^2 = -(r^2 - 1)dt^2 + \frac{1}{(r^2 - 1)}dr^2 + r^2d\phi^2,
\end{equation}
where we have set $r_+ = l = 1$, the general $r_+$ case is a simple generalization. The wave equation $(\Box + m^2)\Phi(t,r,\phi) = 0$ after separation of variables 
$\Phi = e^{-i\omega t}e^{ip\phi}r^{-\frac{1}{2}}\psi_{\omega p}(r)$ leads to a radial equation, which can be expressed in terms of the tortoise coordinate $z$ (defined via $r = \coth{z}$) as
\begin{equation}
    (-\partial_z^2 + V_p(z) - \omega^2)\psi_{\omega p}(r) = 0,
\end{equation}
with $V_p(z)$ is an implicit function of $z$ given by
\begin{equation}
    V_p(z) = (r^2 - 1)\left[\frac{p^2 + \frac{1}{4}}{r^2} + \nu^2 - \frac{1}{4}\right] = \frac{p^2 + \frac{1}{4}}{\cosh^2{z}} + \frac{\nu^2 - \frac{1}{4}}{\sinh^2{z}},
\end{equation}
with $\nu^2 = 1 + m^2$. To solve the radial equation, we introduce a new variable 
\begin{equation}
    y = 1 - \frac{1}{r^2} = \frac{1}{\cosh^2{z}}
\end{equation}
which turns the radial equation into a hypergeometric form and allows solution. The mode normalizable at the boundary and with free wave boundary conditions at the horizon is \cite{FestucciaThesis}
\begin{equation}
    \psi_{\omega p}(r) = C(\omega, p)y^{\frac{i\omega}{2}}(1 - y)^{\frac{\nu}{2} + \frac{1}{4}}\;_2F_1\left(\frac{1 + \nu + i(\omega + p)}{2} ,\frac{1 + \nu + i(\omega - p)}{2} , 1 + \nu; 1-y\right)
\end{equation}
Note that $p\in\mathbb{Z}$ due to the periodicity of $\phi$,
and
\begin{equation}
    C^2(\omega, p) = \frac{4\omega^2}{f(\omega, p)f(-\omega, p)}
\end{equation}
with the Jost function $f(\omega, p)$ given by 
\begin{equation}
    f(\omega, p) = -\frac{2^{i\omega + 1}\Gamma(1 + \nu)\Gamma(1 - i\omega)}{\Gamma(\frac{1 + \nu - i(\omega + p)}{2})\Gamma(\frac{1 + \nu - i(\omega - p)}{2})}.
\end{equation}

The two-sided $G_{12}$ propagator in momentum space is straightforward to compute in the Hartle-Hawking vacuum, see \cite{Vaibhav2} or Appendix A of \cite{FestucciaThesis}. As we mentioned already, this bulk calculation naturally results in a momentum space (in the boundary directions) correlator thanks to the mode expansions, and the result is
\begin{equation}
    \mathcal{G}_{12}(\omega, p;r , r') = e^{-\frac{\beta\omega}{2}}\left[\frac{1}{2\omega}\frac{e^{\beta\omega}}{e^{\beta\omega} - 1}(rr')^{-\frac{1}{2}}\psi_{\omega p}(r)\psi_{\omega p}(r')\right]
\end{equation}
whose explicit form after including the complete form of the radial mode is
\begin{equation}
    \begin{aligned}
        \mathcal{G}_{12}(\omega, p; r,r') = \frac{\Gamma\left[\frac{i\omega + ip + \lambda}{2}\right]\Gamma\left[\frac{-i\omega + ip + \lambda}{2}\right]\Gamma\left[\frac{i\omega - ip + \lambda}{2}\right]\Gamma\left[\frac{-i\omega - ip + \lambda}{2}\right]}{4\pi\Gamma[\lambda]^2}(yy')^{\frac{i\omega}{2}}[(1 - y)(1  -y')]^{\frac{\lambda}{2}} \times\\
        \times \ _2F_1\left[\frac{i\omega + ip + \lambda}{2}, \frac{i\omega - ip + \lambda}{2}, \lambda, 1 - y\right]\;_2F_1\left[\frac{i\omega + ip + \lambda}{2}, \frac{i\omega - ip + \lambda}{2}, \lambda, 1 - y'\right].  \label{momG12}
    \end{aligned}
\end{equation}
Here $y$ is defined as before, $\lambda = 1 + \nu$ and $\beta = 2\pi$. Note that we use $\mathcal{G}$ for momentum space and $G$ for position space.

The extrapolate boundary limit of this propagator is straightforward to compute, and we present it below for completeness. We will not really need it, but it is worthwhile noticing that the Fourier transform simplifies somewhat in this limit \cite{CK}. The boundary correlator in the notation of \cite{FestucciaThesis} is 
\begin{equation}
    \mathcal{G}^{boundary}_{12}(\omega, p) = \lim_{r,r'\rightarrow\infty}(2\nu r^{\lambda})(2\nu r'^{\lambda})\mathcal{G}_{12}(\omega,p;r,r').
\end{equation}
Using \eqref{momG12} in the above prescription, we get 
\begin{equation}
    \mathcal{G}^{boundary}_{12}(\omega, p) = \frac{\Gamma\left[\frac{i\omega + ip + \lambda}{2}\right]\Gamma\left[\frac{-i\omega + ip + \lambda}{2}\right]\Gamma\left[\frac{i\omega - ip + \lambda}{2}\right]\Gamma\left[\frac{-i\omega - ip + \lambda}{2}\right]}{\pi\Gamma[\lambda - 1]^2}.
\end{equation}

\subsection{Position Space}

The bulk BTZ propagator in position space can be obtained as a mirror sum of its covering AdS$_3$ Green function. This is done very clearly in \cite{Ichinose}, whose notations (with $r_+=l=1$) we follow.  The Feynman propagator can be looked up \cite{Ichinose}:
\begin{equation}
    -iG_F^{\text{bulk}}(\Delta t, \Delta\phi; r,r') = \frac{1}{2^{\lambda + 1}\pi}\sum_{n = -\infty}^{\infty}\frac{1}{z_n^\lambda}\;_2F_1\left[\frac{\lambda}{2}, \frac{\lambda + 1}{2}, \lambda; \frac{1}{z_n^2}\right],
\end{equation}
where
\begin{equation}
    z_n = rr'\cosh{(r_+\Delta\phi_n)} - \sqrt{r^2 - 1}\sqrt{r'^2 -1}\cosh{(\Delta t})
\end{equation}
The definitions \eqref{Wightman} and \eqref{G12} together with 
\begin{equation}
    \beta = 2\pi
\end{equation}
allow us to determine (this essentially leads to a $\cosh{(i\pi)}$ factor) 
\begin{equation}
    -iG_{12}^{\text{bulk}}(\Delta t, \Delta\phi; r, r') = \frac{1}{2^{\lambda + 1}\pi}\sum_{n = -\infty}^{\infty}\frac{1}{s_n^\lambda}\;_2F_1\left[\frac{\lambda}{2}, \frac{\lambda + 1}{2}, \lambda; \frac{1}{s_n^2}\right], \label{positionG12}
\end{equation}
where
\begin{equation}
    s_n = rr'\cosh{\left(\Delta\phi_n\right)} + \sqrt{r^2 - 1}\sqrt{r'^2 - 1}\cosh{\left(\Delta t\right)}.
\end{equation}

\subsection{The Transform}

Our goal then is to demonstrate that 
\begin{equation}
    \mathcal{G}_{12}(\omega, p; r,r') = \int_{-\infty}^{\infty}d\Delta t\int_0^{2\pi}d\Delta\phi\;e^{-i\omega\Delta t + ip\Delta\phi}G_{12}^{\text{bulk}}(\Delta t, \Delta\omega; r,r')
\end{equation}
where the LHS is \eqref{momG12} and the RHS contains \eqref{positionG12}. Since the integrand is a sum over $n$ and each $n$ involves a contribution of $\Delta\phi_n = \Delta\phi + 2\pi n$, the integral domain for the new dummy variable $\Delta\phi$ is $(-\infty, +\infty)$. So the actual relation that we wish to demonstrate is 
\begin{equation}
    \mathcal{G}_{12}(\omega, p; r,r')=\int_{-\infty}^{\infty}d\Delta t\int_{-\infty}^{\infty}d\Delta\phi\;e^{-i\omega\Delta t + ip\Delta\phi}\;\frac{1}{2^{\lambda + 1}\pi}\frac{1}{s^\lambda}\;_2F_1\left[\frac{\lambda}{2}, \frac{\lambda + 1}{2}, \lambda; \frac{1}{s^2}\right] \label{RHSintegral}
\end{equation}
where
\begin{equation}
    s = rr'\cosh{\left(\Delta\phi\right)} + \sqrt{r^2 - 1}\sqrt{r'^2 - 1}\cosh{\left(\Delta t\right)}
\end{equation}

\subsection{Tricks}

Inspired by \cite{CK}, we introduce the notation
\bea
   & \Delta\phi = \frac{\phi}{2},\ \Delta t = \frac{t}{2}, \ \ 
    S = m(e^{\frac{\phi}{2}} + e^{-\frac{\phi}{2}}) + n(e^{\frac{t}{2}} + e^{-\frac{t}{2}}), \\ 
    &m = \frac{rr'}{2},\ n = \frac{\sqrt{r^2 - 1}\sqrt{r'^2 - 1}}{2}
\eea
and the new variables
\begin{equation}
    t_1 = \frac{me^{\frac{\phi}{2}}}{S}, t_2 = \frac{me^{-\frac{\phi}{2}}}{S}, t_3 = \frac{ne^{\frac{t}{2}}}{S}, t_4 = \frac{ne^{-\frac{t}{2}}}{S}.
\end{equation}
Thus, the RHS of \eqref{RHSintegral} that we are interested in evaluating becomes
\begin{equation}
    g = \frac{1}{2^{\lambda + 3}\pi}\int_{-\infty}^{+\infty}dt\int_{-\infty}^{+\infty}d\phi\frac{e^{-\frac{i\omega t + ip\phi}{2}}}{S^{\lambda}}\;_2F_1\left[\frac{\lambda}{2}, \frac{\lambda + 1}{2}, \lambda; \frac{1}{S^2}\right] \label{g-1}
\end{equation}
Since they are clearly not all independent, we choose $t_1, t_3$ as the primary coordinates and  with the relations
\begin{equation}
    \frac{t_1t_2}{m^2} = \frac{t_3t_4}{n^2},\  \ t_1 + t_2 + t_3 + t_4 = 1,
\end{equation}
express $t_2, t_4$ and $S$ as
\begin{equation}
    t_2 = \frac{m^2t_3(1 - t_1 - t_3)}{n^2t_1 + m^2t_3}, \ t_4 = \frac{n^2t_3(1 - t_1 - t_3)}{n^2t_1 + m^2t_3}, \ S = \sqrt{\frac{n^2t_1 + m^2t_3}{t_1t_3(1 - t_1 - t_3)}}.
\end{equation}
The Jacobian of the change of variables to $t_1, t_3$ is  
\begin{equation}
    |J| = \frac{2}{t_1t_3(1 - t_1 - t_3)}.
\end{equation}
The integration range in the $t_1$-$t_3$ plane can be determined by noticing that the two quantities below, both span the region $[0,\infty)$:
\begin{equation}
    e^{\frac{\phi + t}{2}} = \frac{n^2t_1 + m^2t_3}{1 - t_1 - t_3}, e^{\frac{\phi - t}{2}} = \frac{nt_1}{mt_3}.
\end{equation}
This translates to the constraint that the region of integration is the triangle defined by
\begin{equation}
    \mathcal{R} = \{t_1, t_3|t_1 >0, t_3 > 0, t_1 + t_3 < 1\}
\end{equation}
and thus we get 
\begin{equation}
    \begin{aligned}
        g(a,b,c)\equiv g = \frac{m^{-a-c-1}n^{-b-c-1}}{2^{a + b + c + 4}\pi}&\int_{\mathcal{R}}dt_1dt_3\frac{t_1^{a}t_3^{b}(n^2t_1 + m^2t_3)^{c}}{(1 - t_1 - t_3)^{1 + c}} \times\\
        &\times\, _2F_1\left[\frac{a + b + 2}{2}, \frac{a + b  + 3}{2}, a + b + 2; \frac{t_1t_3(1 - t_1 - t_3)}{n^2t_1 + m^2t_3}\right] \label{g-2} 
    \end{aligned}
\end{equation}
where
\begin{equation}
    a = \frac{ip + i\omega + \lambda}{2} - 1, b = \frac{-ip - i\omega + \lambda}{2} - 1, c = \frac{ip - i\omega - \lambda}{2} \label{abc}
\end{equation}
The inverted version of these parameters is also useful:
\begin{equation}
    \lambda = a + b + 2, ip = a + c + 1, i\omega = -b - c - 1.
\end{equation}
To ensure that the region of integration is simple, now we make the substitutions
\begin{equation}
    t_1 = uv, t_3 = u(1 - v)
\end{equation}
The benefit here is that the domain of integration for $u,v$ is simply $(0,1)$. Substituting these variables and noting that the Jacobian leads to an extra $u$, we get the integral
\begin{equation}
   \begin{aligned}
        g(a,b,c) = &\frac{m^{-a-c-1}n^{-b-c-1}}{2^{a + b + c + 4}\pi}\int_{0}^1du\int_{0}^{1}dv\;u^{a + b + c + 1}(1 - u)^{-(1 + c)}v^{a}(1 - v)^b \times\\
        &\times (m^2 + (n^2 - m^2)v)^{c} {}_2F_1\left[\frac{a + b + 2}{2}, \frac{a + b  + 3}{2}, a + b + 2; \frac{u(1 - u)v(1 - v)}{m^2 + (n^2 - m^2)v}\right]. \label{guv}
    \end{aligned}
\end{equation}
The momentum space 2-sided correlator \eqref{momG12} in the same notation takes the form
\begin{equation}
    \begin{aligned}
        f(a,b,c) \equiv &\frac{\Gamma[1 + a]\Gamma[1 + b]]\Gamma[-c]\Gamma[a + b + c + 1]]}{4\pi\Gamma[a + b + 2]^2}(yy')^{\frac{-b - c - 1}{2}}[(1 - y)(1  -y')]^{\frac{a + b + 2}{2}}\\
        &\;\;\;\;\;\;_2F_1[1 + a, -c, a + b +2, 1- y]\;_2F_1[1 + a, -c, a + b +2, 1- y']. \label{f}
    \end{aligned}
\end{equation}
So our claim then, is that the integral presented in $g(a,b,c)$ evaluates to $f(a,b,c)$. 

\subsection{Checks}

The integral in $g(a,b,c)$ has close connections to various known integrals in various books on special functions. But unfortunately, though structurally similar, the known integrals we could find were all simpler. 

\begin{figure}
    \centering
    \includegraphics[width=\linewidth]{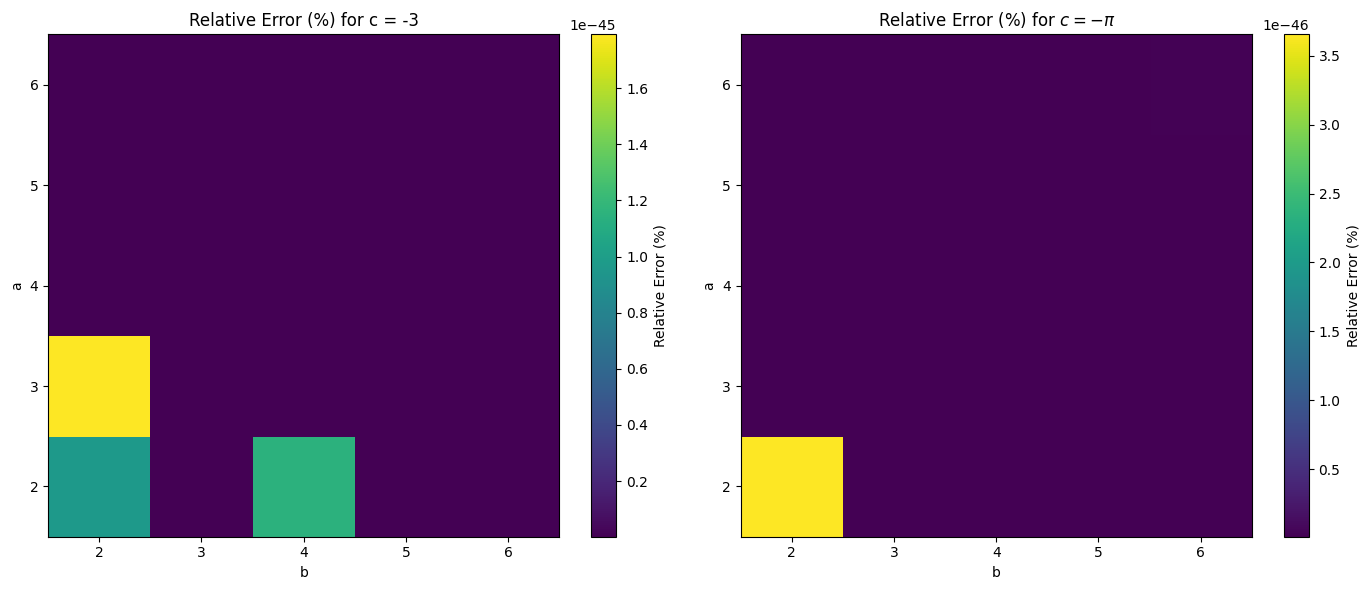}
    \caption{The relative error between the two numerically computed quantities for $c = -3$ and values of $a,b \in [2,6]\times[2,6]$ is on the left. On the right, we have the relative error between the two numerically computed quantities for $c = -\pi$ and values of $a,b \in [2,6]\times[2,6]$. In these plots we only present integer values of $a, b$ within the specified ranges, but the precision is up to 45 decimals. But with precision to 35 decimals instead of 45, we have checked it also for the case $c=-3$ for 10,000 points in the grid $a,b \in [1,10]\times[1,10]$. We have also randomly checked various other values to high precision.}
    \label{fig4}
\end{figure}

But all is not lost, because the structure of the parameters strongly suggests that the integral is analytic in $a, b , c$. We can numerically evaluate it at specific values of these parameters and see whether the left and right sides match\footnote{In fact, this is one of the advantages of the form we reached in \eqref{guv}. Getting the integrand and the range of integration to play well together, is part of the challenge of the Fourier transform. Here, we were able to accomplish that via the analytic continuation in parameters and the compactness of the integration domain. Because the region of integration is compact, it is amenable to high-precision numerical integration. Also, complex phases in parameters often do not help in numerical integrations for reasons loosely analogous to the sign problem in lattice path integrals.}. In particular, we do not have to evaluate it at the complex values of $a, b, c$ that we started with -- we can get considerable confidence in the formula by analytically continuing the parameters to suitable real values, and computing the percentage error between $f$ and $g$. This is what we do in Fig. \ref{fig4}.  Notice that the $r,r'$ behavior is in $y,y',m,n$. One can numerically evaluate the integral for real parameters $(a,b,c)$ for fixed values of $r,r'$.

The plots show that the results are quite good. In Python, we get matches up to 45 decimals. We numerically observe that the following inequalities are necessary for the numerical integrals to converge:
\begin{equation}
    c \le -1, \ \ a + b + c + 1 \ge 0.
\end{equation}
For irrational values of the parameters, the numerics is more time consuming, but the percentage errors are still incredibly small. As a sanity check, we have also computed the percentage error where we compare $f(a,b,c)$ against $g(a,c,b)$ and {\em not} $g(a,b,c)$. We find that the match to the 45th decimal is completely ruined and we instead have a $\sim 90\%$ (i.e., $\mathcal{O}(1)$) percentage error. This is of course the generic expectation when comparing two numbers of the same order of magnitude.

\end{document}